# Pattern-Oriented Analysis and Design (POAD) Theory

Jerry Overton
*Computer Sciences Corporation, CSC*
joverton@csc.com

**Abstract**

*Pattern-Oriented Analysis and Design (POAD) is the practice of building complex software by applying proven designs to specific problem domains. Although a great deal of research and practice has been devoted to formalizing existing design patterns and discovering new ones, there has been relatively little research into methods for combining these patterns into software applications. This is partly because the creation of complex software applications is so expensive. This paper proposes a mathematical model of POAD that may allow future research in pattern-oriented techniques to be performed using less expensive formal techniques rather than expensive, complex software development.*

Keywords: software design patterns; software architecture; software engineering; Pattern-Oriented Analysis and Design (POAD); Kolmogorov complexity

## 1. Introduction

With Pattern-Oriented Analysis and Design (POAD), complex software solutions are created by applying proven design patterns to existing problem domains. By building from proven designs, POAD allows software architects to build complex software faster and with greater quality.

This paper introduces a method for describing pattern-oriented practices using a mathematical language. The goal of this work is to introduce a theory that will allow the examination of abstract patterns in the practice of POAD.

Software is typically much more expensive and time consuming to build and test than new mathematical concepts. Although mathematical models cannot replace research using real software, they may be a relatively inexpensive tool for discovering which new ideas are worth investigating. The mathematical model of POAD developed in this paper may allow future research in pattern-oriented techniques to be conducted by formulating questions as mathematical problems and using calculations to suggest viable solutions (see section 9 for examples).

To be applicable in the actual practice of software engineering, the definitions and predictions of this model must be validated by empirical observation. Yacoub and Ammar [13] describes POAD concepts and methods based on the results of applying pattern-oriented techniques in 4 different case studies. This paper compares formal mathematical definitions to the concepts informally described in Yacoub and Ammar [13] and compares predictions based on calculation to the actual methods used in Yacoub and Ammar [13].

Sections 3 and 4 define the idea of a concept space as a normed, linear space in which each point in the space represents a concept, and distances between points represent the dissimilarity between the corresponding concepts. Section 4 goes on to define software design patterns as functions in concept space. Section 5 uses addition and multiplication to describe the composition of software patterns. Section 6 further extends the theory to include a model of software problems and solutions. Section 7 demonstrates the application of POAD theory to predicting the existence of pattern-oriented techniques and to giving clues about the characteristics of those techniques. Section 8 compares this paper to previous works, and section 9 describes possible future applications.

## 2. Limitations of POAD Theory

Although a mathematical description of POAD may be a powerful tool for discovering quantifiable relationships, it is also a very abstract way of looking at this complex software engineering technique.

To build a mathematical description of POAD, it is necessary to limit the theory only to those features that can be quantified and to look for meaningful relationships between those features. The only features that can be described by such a model are those that can be ascribed numerical values – relative software complexity in the case of this paper.

The result is a theory that can provide a *description* of phenomenon of interest but cannot give an *explanation* of that phenomenon. For example, Equation 10 describes a process for composing patterns in a way that allows the creation of completely new information. However, it does not explain from where this new information comes. In contrast, in the role-based pattern compositions of Riehle [11] and Yacoub and Ammar [13], this additional information is explained to be the result of classes from different patterns participating in more than 1 role.

## 3. Absolute Information

Section 4 introduces a concept space that uses absolute information as a way to measure distances between points in the space. This section gives a formal definition of absolute information.

The amount of information in an entity can be defined by its Kolmogorov complexity: the length of the shortest computer program capable of describing that entity. The concepts in this paper are built using a continuous form of Kolmogorov complexity called second quantized Kolmogorov complexity (SQKC).

SQKC is developed using the notion of a theoretical quantum computer $U$. However, quantum computing is not an essential element of POAD. The introduction of $U$ is necessary only to establish a continuous form of Kolmogorov complexity.

In SQKC, a universal quantum computer $U$ takes in an initial quantum string $\hat{\phi}$ and produces an output quantum string $\hat{\psi}$. The input quantum string $\hat{\phi}$ may be in a superposition of strings of many different lengths:

$$\hat{\phi} = \sum_{i=0}^{\infty} \alpha_i \hat{\imath} \qquad 1$$

The average length $\bar{l}$ of the quantum string $\hat{\phi}$ is the average length of its composites:

$$\bar{l}(\hat{\phi}) = \sum_{i=0}^{\infty} |\alpha_i|^2 l(i) \qquad 2$$

**Definition 1** *SQKC*, denoted $K$, of a quantum string $\hat{\psi}$ is the minimum average length of a program that produces $\hat{\psi}$:

$$K(\hat{\psi}) = \min_{U(\hat{\phi}) = \hat{\psi}} \bar{l}(\hat{\phi}) \qquad 3$$

In cases where $U(\hat{\phi}) = \hat{\psi}$ is empty, SQKC is defined as $\bar{l}(\hat{\psi})$.

The length of a program varies by programming language. However, for any 2 programming languages, there exists a program that can convert a program from the first language into the equivalent program in the second. The difference in program lengths of different languages is invariant up to an additive constant proportional to the size of the conversion program. This invariance makes SQKC independent of any programming language (up to an additive constant), and it is also a suitable measure of the absolute amount of information present in any entity.

SQKC is not computable. It is a theoretical measure used in Section 5 to define the act of pattern composition as the manipulation of information in a given pattern.

**Definition 2** The *Conditional SQKC* of a program $\hat{\psi}$ compared to $\hat{\phi}$ is the minimum SQKC of all programs capable of producing $\hat{\psi}$ given only $\hat{\phi}$:

$$K(\hat{\psi} \mid \hat{\phi}) = \min_{U(\hat{\phi}, \hat{\lambda}) = \hat{\psi}} K(\hat{\lambda}) \qquad 4$$

Conditional SQKC is a measure of the amount of information needed to produce a final state given an initial one.

**Definition 3** The *SQKC Distance* between 2 programs $\hat{\psi}$ and $\hat{\phi}$ is the minimal amount of information needed to translate $\hat{\psi}$ into $\hat{\phi}$ and $\hat{\phi}$ into $\hat{\psi}$:

$$\mu(\hat{\psi}, \hat{\phi}) = K(\hat{\psi} \mid \hat{\phi}) + K(\hat{\phi} \mid \hat{\psi}) \qquad 5$$

The SQKC distance is a measure of all effective similarities between $\hat{\psi}$ and $\hat{\phi}$. It fits an intuitive notion of distance. It is symmetric, positive and obeys the triangle inequality.

## 4. Software Design Patterns

A design pattern describes a problem that frequently occurs in software design along with its proven solution. For example, the Model-View-Controller (MVC) pattern is used to solve the problem of creating flexible graphical user interfaces. With MVC, a Model represents an instance of a domain-specific concept, a View realizes a specific user interface representation of the Model, and a Controller relays inputs from the user to the Model.

The following is a series of definitions that lead to the mathematical definition of a pattern and a pattern's behavior.

**Definition 4** A *concept* is a quantum string $\hat{\phi}$ in a unique superposition with all other concepts such that:

$$\hat{\phi} = \sum_{i=0}^{\infty} \alpha_i \hat{i} \qquad\qquad 6$$

The concept $\hat{\phi}$ has properties analogous to the intuitive notion of a concept where $\alpha_i$ can be interpreted as a measure of how well $\hat{i}$ typifies $\hat{\phi}$.

**Definition 5** A *concept space* $\gamma$ is a normed linear space of concepts where the distance between 2 concepts $\hat{\psi_1}$ and $\hat{\psi_2}$ is $\mu(\hat{\psi_1}, \hat{\psi_2})$.

**Definition 6** A *pattern* $f(\hat{\psi}): \gamma \rightarrow \gamma$ is a function that maps concepts to other concepts. The domain of a pattern is the pattern's *context* and the range is the pattern's *structure*. For example, the context of the MVC pattern is the collection of events received from the user. The structure of the pattern is the relationship maintained between the Model, View and Controller participants.

**Definition 7** The *behavior* of a pattern $f$ is its derivative $f'$.

The derivative of a function is itself a function. This new function describes changes in the original function at any point. Similarly, the behavior of a pattern is itself a pattern. This new pattern describes how the original pattern changes given a particular context.

The behavior of the MVC pattern, for example, describes how events from the user are processed by the controller, how the controller updates the model, and how the views respond to changes in the model.

## 5. Composition of Design Patterns

Simpler patterns can be composed into more complex patterns. For example, the MVC pattern is the result of multiplication of the Observer and Strategy patterns. A model class is introduced that implements both the Observer's subject class and the Strategy's strategy class. The resulting MVC pattern has qualities (such as dynamic user event handling) found in neither original pattern.

The POAD practice described in Yacoub and Ammar [13] relies on 2 basic pattern operations: stringing patterns and overlapping patterns. Two patterns are strung together by combining the participants of the patterns. For example, stringing together a Strategy pattern and an Observer pattern, will result in a pattern that has all the classes of the Strategy pattern and all the classes of the Observer pattern.

When stringing together patterns, the resulting composite pattern is never more complicated than the sum of the originals. Stringing together patterns is commutative and associative: no matter what order a series of patterns are strung together, the result will be the same. Any pattern strung together with the null pattern would result in the original pattern. Any pattern strung together with its anti-pattern would result in the null pattern.

With pattern overlapping, a single class participant from a pattern is also made to be a class participant of other patterns in a single design. For example, one way of overlapping the Observer pattern with the Strategy pattern is to use all participants of both patterns, but make the Abstract Strategy class of the strategy pattern also play the role of Abstract Subject in the Observer pattern.

Overlapping patterns can produce complexity not present in the original patterns. Any pattern overlapped with a pattern consisting of a single, empty class will result in the original pattern. Changing the order of overlapping a series of patterns does not change the outcome. A particular pattern overlap can be performed on individual patterns before they are strung together or on a composite after patterns are strung together with the same result.

The following introduces a pattern algebra that models, mathematically, the composition of design patterns described in Yacoub and Ammar [13]. In this model, pattern addition is an abstract model of stringing together patterns, and pattern multiplication is an abstract model of overlapping patterns.

**Definition 8** The *pattern distance* $\|f-g\|$ between 2 patterns $f$ and $g$ on the interval $[\hat{\psi_a}, \hat{\psi_b}]$ is:

$$\int \mu(f(\hat{\phi}), g(\hat{\phi})) \, d\hat{\phi} \qquad\qquad 7$$

Pattern distance measures the amount of dissimilarity between patterns. The distance between 2 patterns is the total amount of dissimilarity between the patterns when compared using a common context.

**Definition 9** A *pattern space* $\wp$ is a normed linear space of patterns where the distance between 2 patterns $f$ and $g$ is $\|f-g\|$. From Definition 8, the norm $\|f\| = \|f-0\|$ for $\wp$ is a numerical function that satisfies:

a) $\|f\| \geq 0$
b) $\|f\| = 0 \Leftrightarrow f = 0$
c) $\|\lambda f\| = |\lambda| \cdot \|f\|$  (if $\lambda \in \mathbb{R}$)
d) $\|f + g\| \leq \|f\| + \|g\|$

                        8

In $\wp$, the pattern operations of addition and multiplication of patterns $f$, $g$ and $h$ and real numbers $\alpha$ and $\beta$ are defined as:

*Pattern addition*:

a) $f + g = g + f$
b) $(f + g) + h = f + (g + h)$
c) $\forall f \in \wp$, $f + 0 = f$
d) for each $f \in \wp$, $f + (-f) = 0$

                        9

*Pattern multiplication:*

a) $\forall f \in \wp$, $f \cdot 1 = f$
b) $\alpha(\beta f) = (\alpha \beta) f$
c) $(\alpha + \beta) f = \alpha f + \beta f$
d) $\alpha(f + g) = \alpha f + \alpha g$

                        10

The norm of a pattern turns out to be a measure of the amount of information in that pattern. From definitions 3, 9 and 8 it follows that:

$$\begin{aligned}\|f\| &= \|f - 0\| \\ &= \int \mu(f(\widehat{\phi}), 0) \, d\phi \\ &= \int K(f(\widehat{\phi}) \mid 0) + K(0 \mid f(\widehat{\phi})) \, d\phi \\ &= \int K(f(\widehat{\phi})) \, d\phi\end{aligned}$$

                        11

For pattern addition: $\|P_1 + P_2\| \leq \|P_1\| + \|P_2\|$. This places a limit on the amount of complexity that can be produced as a result of an addition operation. With pattern addition, the result can never be more complex than the sum of the original patterns.

For pattern multiplication: $\|\lambda P\| = |\lambda| \cdot \|P\|$. Unlike pattern addition, the amount of complexity produced is not by to the complexity of the original pattern. It is possible to create new information as a result of pattern multiplication.

## 6. Software Problems

This section extends POAD theory to include a model of software problems and patterns solutions.

**Definition 10** A *software problem* is a functional (a function of functions) $J[y_1, \ldots, y_n]$ of the patterns $y_1(\hat{x}), \ldots, y_n(\hat{x})$ of the form:

$$J[y_1, \ldots, y_n] = \int_b^a F(\hat{x}, y_1, \ldots, y_n) \, d\hat{x} \qquad 12$$

Where its *solution* is $J[s_1, \ldots, s_n]$ such that $J$ is an extrema. Solving the problem means finding the patterns that result in the optimal solution: the fastest completion time, the least amount of resources, the best user experience, etc.

## 7. Application of POAD Theory

This section demonstrates the application of POAD theory to predicting the existence of pattern-oriented techniques and to giving clues about the characteristics of those techniques. The predictions made by the theory are compared to conclusions reached in actual practice.

Using the results of 4 case studies, Yacoub and Ammar [13] concluded POAD is a viable application development technique. It is possible to build software applications entirely by stringing together and overlapping design patterns. Using POAD Theory, this same conclusion can now be derived mathematically.

In terms of POAD Theory, the question becomes: Is it possible to approximate the solution $S = J[s_1, \ldots, s_n]$ to an arbitrary software problem $J$, with a pattern composition $P = \alpha_1 p_1 + \ldots + \alpha_n p_n$ that minimizes:

$$\max_{1 \leq i \leq m} \left\| \sum \alpha_i p_i(\hat{x}_i) - S(\hat{x}_i) \right\| \qquad 13$$

In other words, is it possible to find a pattern $P$ that best approximates the solution $S$, given that the only things that are known about $S$ are $S(\hat{x}_i), \ldots, S(\hat{x}_n)$?

Answering this question is the same as asking if there is a solution to the linear system of equations:

$$\begin{bmatrix} p_1(\hat{x}_1) + \cdots + p_n(\hat{x}_1) \\ \vdots \\ p_1(\hat{x}_n) + \cdots + p_n(\hat{x}_n) \end{bmatrix} \begin{bmatrix} \alpha_1 \\ \vdots \\ \alpha_n \end{bmatrix} = \begin{bmatrix} S(\hat{x}_1) \\ \vdots \\ S(\hat{x}_n) \end{bmatrix} \qquad 14$$

Equation 14 is exactly determined and, therefore, has a solution. For every problem $J$, there is a pattern composition $P = \alpha_1 p_1 + \ldots + \alpha_n p_n$ that best approximates the solution $S$ to $J$ given that $S(\hat{x}_i), \ldots, S(\hat{x}_n)$ is known.

POAD Theory can be also be used to give clues about the qualities of a successful pattern-oriented technique. For example, the dependencies between the patterns $s_1(\hat{x}), \ldots, s_n(\hat{x})$ that solve the problem $J$ can be represented as a system of simultaneous of differential equations:

$$s_1' = f_1(s_1, \ldots, s_n)$$
$$\ldots$$
$$s_n' = f_n(s_1, \ldots, s_n)$$

15

Equation 15 assumes that any change in a pattern is some function of all other patterns, and changes in any one pattern has some effect on all the others.

However, one of the conditions necessary for the patterns $s_1(\hat{x}), \ldots, s_n(\hat{x})$ to solve $J$ is that they satisfy a condition:

$$s_i' = z_i$$

16

of the canonical form of the Euler equations, where $z_i$ is a function independent of all other variables of Equation 15. Under this condition, the patterns $s_1, \ldots, s_n$ are eliminated from the functions $f_1, \ldots, f_n$ leaving patterns that are completely independent of each other. Every pattern-oriented solution $J[s_1, \ldots, s_n]$, in the simplest form, consists of patterns $s_1, \ldots, s_n$ that are functionally independent of each other.

Yacoub and Ammar [13] reached this same conclusion by applying pattern-oriented techniques in 4 different case studies. They recommend that solutions be documented as a collection of constructional design patterns. These patterns interact with each other only through well-defined interfaces that allow all patterns to be treated as black boxes with hidden internal structures.

## 8. Related Work

Yacoub and Ammar [13] use several case studies to explore the viability of actual pattern-oriented techniques. Their work is used as an empirical basis for the mathematical model developed in this paper. This paper compares its formal mathematical definitions to their informal concepts and compares predictions based on calculation to the actual methods they used.

Patterns in this paper match abstract models of lattices described by Eden, Yehudai and Gil [6]. A lattice is a meta programming tool that, similar to this paper's definition of a pattern, produces a solution structure given an input context. Unlike the model of patterns in this paper, lattices also include a detailed description of the semantics of how the conversion from context to solution occurs. These semantics included specific techniques called tricks. Although tricks may be used for pattern composition, Eden, Yehudai and Gil [6] focused mainly on using tricks to implement a single pattern and did not explore the possibility or limitations of using tricks for pattern composition.

Patterns in this paper also match abstract models of the patterns described by Hallstrom and Soundarajan [9]. Hallstrom and Soundarajan [9] define patterns as a construct that produces a solution structure (called a reward), given an input context (called a responsibility). Unlike this paper, Hallstrom and Soundarajan [9] focus on reasoning about patterns themselves rather than exploring pattern-oriented techniques. Their formal specifications are designed to identify the standard and specialized portions of a pattern, discover restrictions on the application of a pattern, and predict the rewards of applying a pattern.

Riehle [11] presents examples of how to create composite design patterns using a role-based analysis. Required collaborations are grouped and assigned to roles. Pattern composition occurs when objects from different patterns are assigned to roles based on composition constraints associated with the role. The model of pattern composition presented in this paper works as an abstract model of the role-based composition approach of Riehle [11]. This makes sense because the pattern composition of Yacoub and Ammar [13] is based on the approach described in Riehle [11].

## 9. Future Applications

With further development, it may be possible to use POAD Theory to discover completely new techniques. For example, Section 7 concluded that it was possible to compose software solutions completely from software design patterns. However, it also requires that points of the solution $S(\hat{x}_i), \ldots, S(\hat{x}_n)$ be known a priori. Are there methods that allow the composition of software solutions using patterns while requiring a limited knowledge of the solution?

Is it possible to approximate the solution $S = J[s_1, \ldots, s_n]$ to the software problem $J$ by creating a pattern composition $P=f(p_1,\ldots,p_n)$ in iterations knowing only $S(\hat{x}_i)$ in the $i^{th}$ step? If so, how many iterations are needed to approximate the solution to desired accuracy?

A serial approach to defining detailed requirements early in a software development project often leads to a significant amount of wasted effort, according to Ambler [1]. Answers to the questions about what could be done with limited knowledge of the solution may lead to new techniques that allow complex software problems to be solved without requiring detailed upfront specifications of the solution.

Missing from POAD theory is the notion of a software quality. Software qualities are attributes of design patterns such as reliability, usability and performance. Any notion of a software quality $Q$ would most likely be expressed as some function $Q(P)$ of a pattern $P$. Are there methods of composing a pattern $P=f(p_1,\ldots,p_n)$ that can guarantee that the final composite pattern $P$ will preserve the qualities of its composite elements: $Q_i(P)=Q_i(p_i)$ ?

The ability to predict the qualities of complex software prior to its construction is a holy grail of software engineering, according to Bass, Clements, and Kazman [2]. An affirmative answer to the questions about predicting software qualities could lead to techniques that would allow the prediction of even the most complex system's qualities prior to construction. A negative answer may at least indicate that it is not wise to spend time and effort pursuing this goal.